\documentclass[12pt,preprint]{aastex}
\usepackage{graphicx}

\begin{document}

\title{The Rest-frame Ultraviolet Light Profile Shapes of Ly$\alpha$--Emitting Galaxies 
at $z=3.1$\footnote{Based on
    observations made with the NASA/ESA Hubble Space Telescope, and
    obtained from the Hubble Legacy Archive, which is a collaboration
    between the Space Telescope Science Institute (STScI/NASA), the
    Space Telescope European Coordinating Facility (ST-ECF/ESA) and
    the Canadian Astronomy Data Centre (CADC/NRC/CSA). }}

\author{Caryl Gronwall\altaffilmark{2}, Nicholas A. Bond\altaffilmark{3}, Robin Ciardullo\altaffilmark{2}, Eric Gawiser\altaffilmark{3}, Martin Altmann\altaffilmark{4}, 
Guillermo A. Blanc\altaffilmark{5}, John J. Feldmeier\altaffilmark{6} }
\affil{\altaffilmark{2}Department of Astronomy and Astrophysics, Pennsylvania State University, University Park, PA 16802} 
\affil{\altaffilmark{3}Physics \& Astronomy Department, Rutgers University
    Piscataway, NJ 08854}
\affil{\altaffilmark{4}University of Heidelberg, Center for Astronomy, M\"onchhofstr. 12-14, D-69120 Heidelberg, Germany }
\affil{\altaffilmark{5} Department of Astronomy, University of Texas at Austin, Austin, TX 78712}
\affil{\altaffilmark{6} Department of Physics and Astronomy, Youngstown State University, Ohio 44555}

\email{caryl@astro.psu.edu}

\begin{abstract}

We present a morphological analysis of the rest-frame ultraviolet emission 
of 78 resolved, high signal-to-noise $z \sim  3.1$ Lyman Alpha Emitters (LAEs) in the 
Extended Chandra Deep Field South (ECDF-S).  Using {\it HST/ACS\/} $V$-band images taken 
as part of the Galaxy Evolution from Morphology and SEDS (GEMS), Great Observatories 
Origins Deep Survey (GOODS), and Hubble Ultra Deep Field (HUDF) surveys,  we investigate 
both single component and multi-component LAEs, and derive concentration indices,
S\'ersic indices, ellipticities, and half-light radii for all resolved components and systems with a signal-to-noise 
$S/N > 30$.  We show that, although the LAE population is heterogeneous is nature, 
most Ly$\alpha$ emitters highly concentrated, with a distribution of $C$ values 
similar to that measured for field stars; this suggests that the diagnostic is a poor 
discriminator near the resolution limit.  The LAEs also display a wide range of
S\'ersic indices ($0 < n < 12$), similar to that seen for galaxies in the local neighborhood.
However, the majority of LAEs have $n < 2$, and a visual inspection of the images
suggests that the small$-n$ objects have extended or multimodal luminosity profiles, 
while the LAEs with $n > 2$ have compact components surrounded by diffuse emission. 
Moreover, unlike nearby spiral galaxies, whose distribution of ellipticities is flat,
the LAE ellipticity distribution peaks near $1 - b/a \sim 0.55$.   Thus, the population
has more in common with $z \sim 3$ Lyman-break galaxies than local star-forming
objects.     

\end{abstract}

  \keywords{cosmology: observations -- galaxies: formation -- galaxies: 
  high-redshift -- galaxies: structure}

\vspace{0.4in}

\section{Introduction}

In the local universe, most galaxies fall onto the Hubble Sequence
\citep{HubbleSeq} which runs from older, red, quiescent galaxies with
a $r^{1/4}$ surface-brightness profiles to star-forming, gas-rich disks
with exponential luminosity distributions.  These light profiles are generally
described by the \citet{Sersic} parameter,  defined through $I(r) \propto r^{1/n}$,
where $n$ typically ranges from $\sim 1$ (an exponential disk) to
$\sim 4$ (an r$^{1/4}$-law spheroid).  
Out to intermediate ($z \sim 1.5$) redshifts, galaxies can generally be placed
onto the local Hubble Sequence quite reliably \citep{Cons04}.  However,
at larger distances, this sequence breaks down, as most sources appear clumpy and 
irregular \citep[e.g.,][]{Steidel96,Papovich05,Conselice05,Venemans05,Pirzkal07}
and are therefore difficult to classify.   Consequently, although the morphological 
parameters of distant galaxies can be quantified via the use of codes such as 
{\small GALFIT} \citep{GALFIT} and {\small GIM2D} \citep{GIM2D},
these parameters do not necessarily directly translate into the familiar Hubble sequence
seen in the nearby universe.
Moreover, the robustness of the derived parameters is highly dependent on the 
signal-to-noise ($S/N$) of the measurement,  with simulations
suggesting that a $S/N$ of at least 15 is required \citep{Ravindranath06} to 
reliably fit a high-redshift galaxy.

To address some of these difficulties, a number of non-parametric
measures of galaxy morphology have been developed,
including the concentration, asymmetry, and clumpiness indices
\citep[CAS;][]{CAS}, and the Gini coefficient, which measures a
profile's departure from uniformity \citep{Lotz04}.  These systems have been
used extensively on samples of high-redshift galaxies, since they 
require no a priori knowledge about the functional form of the luminosity distribution.  
In addition, unlike other fitting techniques which use smooth two-dimensional
profiles, these non-parametric techniques can also quantify the non-uniformity of a
galaxy (i.e., test for the presence of star-forming regions), and define its asymmetry.
Their one drawback, of course, is that the values one obtains may depend on the
image depth, and that this sensitivity has not been well-quantified for 
high-redshift galaxies.

The majority of galaxies known at high ($z \geq 2.5$) redshifts have been identified
via the Lyman-break technique, wherein galaxies with Lyman-limit discontinuities are 
identified by their unique location in color-color space \citep{Steidel96}.
Morphological analyses of these Lyman Break Galaxies (LBGs) in the rest-frame 
UV have revealed that most of these systems are disturbed and disk-like (i.e., with 
exponential profiles), and only $\sim 30\%$ have profiles consistent with galactic
spheroids \citep[e.g.,][]{Ferguson04,Lotz06,Ravindranath06}.
Moreover, \citet{Ravindranath06} have found that the distribution of
ellipticities derived from GALFIT are skewed towards higher
values, and that this asymmetry becomes more pronounced at
higher redshift.  
This dominance of ``elongated'' morphologies seen in LBGs has been
interpreted as evidence for galaxy formation within filamentary structure or
the presence of proto-bars that span the entire visible disk of the galaxy.

Ly$\alpha$ Emitters(LAEs) are a class of high-redshift galaxies
which have not been studied as extensively as LBGs.   Like their
Lyman break counterparts, LAEs between $2 < z < 4$  are 
actively star-forming objects \citep[e.g.,][]{CowieHu}, and at the bright end,
the sample of LAEs overlaps that of LBGs.   However, a typical Ly$\alpha$
emitting galaxy has  a lower stellar mass ($\sim 10^9 M_{\sun}$),  higher
mass-specific star-formation rate ($\sim 10^{-8} M_{\sun}$~yr$^{-1} M_{\sun}^{-1}$, 
and lower dust content ($E(B-V) \sim 0.1$) \citep{Venemans05,Gawiser07,lai}
than any galaxy discovered via the Lyman-break technique.  Moreover, 
LAEs are less clustered than LBGs, suggesting a lower dark-matter halo
mass, and their bias factor $b \sim 1.7$) consistent with that expected from the
progenitors of present-day $L^*$ galaxies \citep{Gawiser07,guaita}.  


To date, relatively few LAEs have been studied morphologically.  In the
rest-frame UV, most $3 \lesssim z \lesssim 6$ LAEs are small ($\lesssim 1$~kpc),
compact ($C>2.5$), and barely resolved at {\it HST\/} resolution
\citep{Venemans05,Pirzkal07,Overzier08,taniguchi}.  Those LAEs that 
are not compact  ($\sim 20 - 45$\%) exhibit clumpy or irregular
morphology, with diffuse components extending to several kiloparsecs.  
Due to their small sizes and low luminosities, however, categorizing these
objects within the context of existing classification schemes has been difficult.


Here we present a study of the rest-frame ultraviolet morphology of a
statistically-complete sample of $z=3.1$ LAEs \citep{GronwallLAE} present on
archival {\it HST}/ACS\/ $V$-band images from the 
Galaxy Evolution from Morphology and SEDS
\citep[GEMS,][]{GEMS}, Great Observatories Origins Deep Survey
\citep[GOODS,][]{GOODS}, and Hubble Ultradeep Field
\citep[HUDF,][]{HUDF} surveys.  
In Paper I \citep{Bond09}, we investigated the size distribution of LAEs by 
developing an analysis pipeline which identifies each component of an LAE, 
and measures its size and brightness.  We found that most LAEs are extremely
compact, with half-light radii, $r_e < 2$~kpc, and that a $S/N > 30$ is required
for a robust measure of morphology.  Here, we extend this study, and provide
more detailed morphological information on each object.   In \S~\ref{sec:data} 
and \ref{sec:method}, we describe the data and detail the algorithms used in 
our morphology-fitting pipeline analysis.   In \S~\ref{sec:results}, we present the 
best-fitting morphological parameters for each galaxy, and explore how these 
parameters vary with image depth.  Finally, in \S~\ref{sec:discussion}, we discuss 
the implications of our findings.  Throughout this paper, we will assume a
concordance cosmology with $H_0=71$~km~s$^{-1}$~Mpc$^{-1}$,
$\Omega_m=0.27$, and $\Omega_{\lambda}=0.73$ \citep{WMAP}.  With these
values, $1\arcsec = 7.75$~physical~kpc at $z=3.1$.


\section{Data}
\label{sec:data}

Our study uses the statistically complete sample of $z = 3.1$ LAEs
identified by \citet{GronwallLAE} in their narrow-band 4990~\AA\ survey of the
Extended Chandra Deep Field-South; these objects are defined to have emission-line  
fluxes $F_{4990} > 1.5 \times 10^{-17}$~ergs~cm$^{-2}$~s$^{-1}$ and
observed-frame Ly$\alpha$ equivalent widths EW$>80$~\AA\null.  As
explained in Paper I, the sample contains $154$ LAEs in a 0.32~deg$^2$
area of sky, and $116$ of the objects fall in fields observed by {\sl HST}.  
A full description of the data is given in Paper I  (see Table I in Paper I), but
to summarize:  the GEMS survey includes $97$ LAEs and reaches a 
$V$-band $5 \sigma$ point-source depth of $28.3$ mag, the GOODS survey (Version 2.0)
contains $29$ LAEs and goes to a point-source depth of $V  = 28.8$,  and HUDF 
contains $3$ LAEs and reaches a depth of $V = 30.2$.  In addition, the GEMS survey
includes the first epoch of GOODS observations (reduced using their pipeline, hereafter sGOODS), and 
this sGOODS data reaches a depth of $V = 27.9$.  We note that 22/29 of our
GOODS LAEs are also covered by sGOODS, 9 of the GOODS LAEs are also in
GEMS, and all 3 LAEs in the HUDF are also part of GOODS.  This overlaps
allows us to test the depth-dependence of some of our morphological diagnostics.

We retrieved $V_{606}$-band ACS images from the Multimission Archive at the Space Telescope Science Institute.  We chose $V$ for our study, in order to use the GEMS frames (which
are $V$-band only).
Additionally, in $i$ or $z$,  most of our GOODS LAEs are either undetected or 
or too faint for a robust analysis.
Although $V_{606}$  does include the flux from the
Ly$\alpha$ emission line, our ground-based narrow-band measurements demonstrate that  
this line-emission accounts for only 10 to 20\% of the total counts detected through the 
broad $V$ filter.     (59\% of our LAEs have a contamination fraction of less than 10\% and 92\% of
the sample has a contamination fraction less than 20\%.)
Thus, our images, are indeed primarily recording the rest-frame UV continuum
generated by stars, rather than the line-emission from the gas.


Since the raw ACS frames undersample the point spread function (PSF),  GEMS,
GOODS, and the HUDF all  employed a multidrizzle dithering technique
\citep{Multidrizzle}, which reduced the pixel scale from $0\farcs 05$
to $0\farcs03$.   For GEMS and GOODS, this procedure resulted in correlated noise
between the pixels, thereby complicating the interpretation of the $\chi^2$ values generated
by profile-fitting algorithms (see \S~\ref{sec:method}).  The HUDF images, however,
have no correlated noise, as the number of exposures was large enough
for sub-samples to be combined independently \citep[i.e., by setting {\tt
PIXFRAC}$=1$ in ``Multidrizzle'',][]{Multidrizzle}.

\section{Methodology}
\label{sec:method}

The full morphology pipeline described in Paper I, works in five stages: cutout extraction from
survey images, source detection \citep[using SExtractor;][]{SExtractor}, centroid estimation 
and aperture photometry (using {\tt PHOT}), light profile fitting \citep[using GALFIT;][]{GALFIT},
and point sources identification.  Our cutouts were defined as
$80 \times 80$ pixel ($2\farcs4 \times 2\farcs4$) regions around each object; this area was
deemed large enough area to compensate astrometric errors in the LAE positions, and 
allow for photometry of components located on the edge of our selection region.
SExtractor was then used on each cutout, to identify all sources consisting of 9 contiguous 
pixels above a1.65$\sigma$ detection threshold.  As described in Paper I,  each source
located within $0 \farcs 6$ of the ground-based Ly$\alpha$ position was defined as an
LAE component, and the center of the LAE system was defined as the flux-weighted mean 
position of all detections within this selection radius.  At the same time, we also used 
SExtractor to fit and subtract a uniform sky from the cutout, in order to not remove any 
diffuse Ly$\alpha$ emission; in the local universe, such emission can extend several 
half-light radii from the center of a galaxy \citep{ostlin09}.   

To compute the photometric centroid of each LAE,  we again used SExtractor, this time
identifying all objects 
with 5 contiguous pixels above the $1.65 \sigma$ threshold and computing their
flux-weighted mean position.  This enabled us to include components too dim or small for 
morphological measurements.   Using this position, we then measured each LAE's total flux 
and half-light radius, under the assumption that the total LAE flux is contained with our
$0\farcs 6$ selection radius.   This should be a good assumption, as LAEs are typically 
quite small ($<1$ kpc)  in the rest-frame UV \citep{Venemans05,Pirzkal07,Overzier08,taniguchi}
and many of our objects  remain unresolved even at the depth of the current HST images.

Finally, as described in Paper I, we use GALFIT \citep{GALFIT} to simultaneously fit each 
detection to a S\'{e}rsic profile \citep{Sersic}
\begin{equation}
I(r) = I(0) \exp \left\{-b \left[ \left( {r \over r_e} \right)^{1/n}- 1 \right] \right\},
\label{eq:Sersic}
\end{equation}
where $r_e$ is the half-light radius, the parameter $n$ characterizes the steepness of 
the light profile, and $r$ represents the radius vector of model isophotes, i.e.,
\begin{equation}
r=\left(\left|x-x_c\right|^2+\left|\frac{y-y_c}{q}\right|^2\right)^{1/2},
\label{eq:ellipse}
\end{equation}
with $x_c$ and $y_c$ being the model's centroid position, and $q$ being its axis
ratio ($b/a$).   To facilitate the computation, we generally initialized the S\'ersic
indices to $n = 4$, i.e., a \citet{deV} profile, rather than a $n = 1$ exponential disk, 
and then allowed SExtractor to provide starting values for the object's position, axis ratio, 
position angle, total magnitude, $V_{GF}$, and half-light radius, $r_{e,GF}$\footnote{We will 
refer to the best-fit value of this parameter as $r_{e,GF}$ to distinguish it from other measures 
of the half-light radius.}.  In addition, to prevent the program from finding unrealistic solutions, we 
placed constraints on several of the parameters; specifically, $24.0 < V_{GF} < 29$, 
$0.1 < r_{e,GF} < 15$~pixels, $0.1 < n < 15$, $0.1 < q < 1$, $\Delta x < 2$, and
$\Delta y < 2$.  As in Paper I, we define an LAE component as unresolved if the reduced 
$\chi^2$ of a fit of its surface-brightness profile using the {\sl HST\/} PSF is smaller than that 
of a S\'{e}rsic profile.
 
 
Note that {\small GALFIT's} $\chi^2$ minimizations were performed using
uncertainties provided by the weight-map images created by 
multridrizzle (see \S~\ref{sec:data}).  However, because the
drizzling process often maps a single pixel of an input image
onto multiple pixels of the output image, the noise properties
of adjacent pixels are not necessarily independent \citep{casertano}. 
Since this co-variance is not included in the weight map, the 
$\chi^2$ values found by {\small GALFIT} on the GEMS and GOODS images
are significantly less than one.  To correct for the correlated noise, we  
reduce the degrees of freedom in the fit by a factor related to the pixel subsampling.
We computed the correction factors within 40-pixel cutouts by computing the 3-sigma-clipped pixel-to-pixel variance of flux values and comparing it to the median variance given by the weight images.
For the GOODS data, this was empirically estimated to be $f_{GOODS}=2.46$;
for GEMS, $f_{GEMS}=1.96$. 

Unless otherwise specified, our fitting procedure was applied to every object
within each cutout.  However, we only report the properties of those components
that fall within the LAE selection circle.  No bad pixel masks were used, but
each fit was inspected by eye; when problems were encountered, such as
bad fits, or warning errors in SExtractor, they were treated on a
case-by-case basis (see \S~\ref{sec:results}).


\section{Results}
\label{sec:results}


As discussed in Paper I, SExtractor was used to detect and compute the total magnitudes
(MAG\_AUTO) of each individual LAE component within our $0 \farcs 6$ selection radius.  
In Paper I, we found that of the 97 LAEs covered by the GEMS survey, 21 (22\%) have no counterpart in the {\sl HST\/} images.  Of the remaining objects, 76 (78\%)  have at least one 
component detected within the $0 \farcs 6$ selection circle, 16 (16\%) have at least 2 
components, and 4 (4\%) have at least 3 components.   Similarly, of the 29 LAEs covered by 
the GOODS survey, 6 (21\%) have no counterpart on the HST frames, while 4 (14\%) have two
components, and one (3\%) is complex (with five components).  Interestingly, all three
LAEs observed in the HUDF are identified as single component systems with asymmetric, 
extended emission.  However, this morphology, which is observable on both the GOODS and the 
HUDF images, is probably not representative of the overall LAE population.  In fact, as shown 
by \citet{Bond09}, 50\% of the \citet{GronwallLAE} LAEs are unresolved even at GOODS depth.

We note that our ability to detect multiple components is sensitive to survey depth.  This is
evident from a comparison of our results from sGOODS subset of GOODS:  while 12 of the 17 
individual components are unresolved in sGOODS images only 4 of these12 ``point sources''
remain unresolved at the GOODS depth.  It is therefore likely that a deeper imaging surveys 
would resolve more of the LAE components, and reveal extended emission that is undetectable
with current data.  In fact,  the majority of the multiple rest-frame UV ``clumps" detected on the 
{\sl HST\/}frames are probably individual star-forming regions within a single, larger system, or 
possibly the result of an ongoing merger. 

For the analysis presented below, we analyze the morphology of each component 
individually as well as the LAE system as a whole.  Like the LAE systems,  approximately 
half of the observed LAE components are unresolved:  this includes 
50 of the 95 components in GEMS and 15 of the 31 components in GOODS. 
 
 Tests performed in Paper~I suggest that a $S/N > 30$ within a fixed half-light radius is 
 required to robustly determine the size of a galaxy.  Thus,  in this paper, we measure the concentration index and present the results of our GALFIT fits for ellipticites, S\'ersic profiles, and half-light radii for each {\it resolved\/} component with $S/N > 30$, as well as
for each LAE system with $S/N>30$.  This signal-to-noise cut 
corresponds to $V_{GF} <28.5$ for HUDF, $V_{GF} <26.8$ for GOODS, and $V_{GF} < 26.5$ for 
GEMS.

\subsection{Concentration}
\label{subsec:concentration}
The Concentraton, Asymmetry, ClumpinesS system, or CAS \citep{CAS} was developed
to estimate the morphology of distant galaxies quantitatively.  
We did not make extensive tests of measuring clumpiness for our sample as the majority of the sample
did not show multiple clumps on visual inspection and those that did only showed a few clumps at most.
We were also concerned that the low surface brightness and  small angular extent of the LAEs
would preclude an accurate measurement of clumpiness.
We note that \citep{Pirzkal07} did not attempt measuring clumpiness for their sample while they
did measure concentration and asymmetry.
We did attempt to measure asymmetry for our sample and found that the asymmetry parameter routinely
was derived to be quite small ($A \lesssim 0.2$) even for objects that are visually extended and asymmetric.  We do not
believe these results to be representative  of the asymmetry of our sample.
We also found that the results for asymmetry were dependent on the depth of the images used. 
It is possible that a deeper survey at HST resolution may be able to provide more robust measurements 
of these paramters, but it is beyond the scope of this paper to 
determine the requirements for such a study.  

We could, however, determine the Concentration indices of our LAE sample.  
Following \citet{CAS}, we define this index via
\begin{equation}
C = 5 \times \log (r_{80\%} / r_{20\%})
\end{equation}
where $r_{80\%}$ and $r_{20\%}$ are the circular radii containing 80\% and 20\% of the
total flux.  Table~\ref{tab:Morph} lists these values for each resolved LAE component,
and Table~\ref{tab:systems} gives the measurements for LAE systems.   Entries missing
from the table represent are for extremely faint sources, for which we were unable to
calculate a value.   

With the exception of one object (LAE 29),  the concentration indices for all the individual 
components  lie in the range $2 < C < 3.5$, with a median value of $\sim 2.5$; for the
multi-component systems, which generally contain additional extended light,
$1.9 < C < 3.8$ with a median of $C \sim 2.6$.  For comparison, elliptical galaxies
and spiral bulges in the local universe have values of $C \gtrsim 4$, while in nearby
disk systems $3 < C < 4$.  Concentration indices as low as $C \sim 2.5$ are generally 
associated with low-surface brightness and irregular galaxies.

\begin{figure}[t]
\plotone{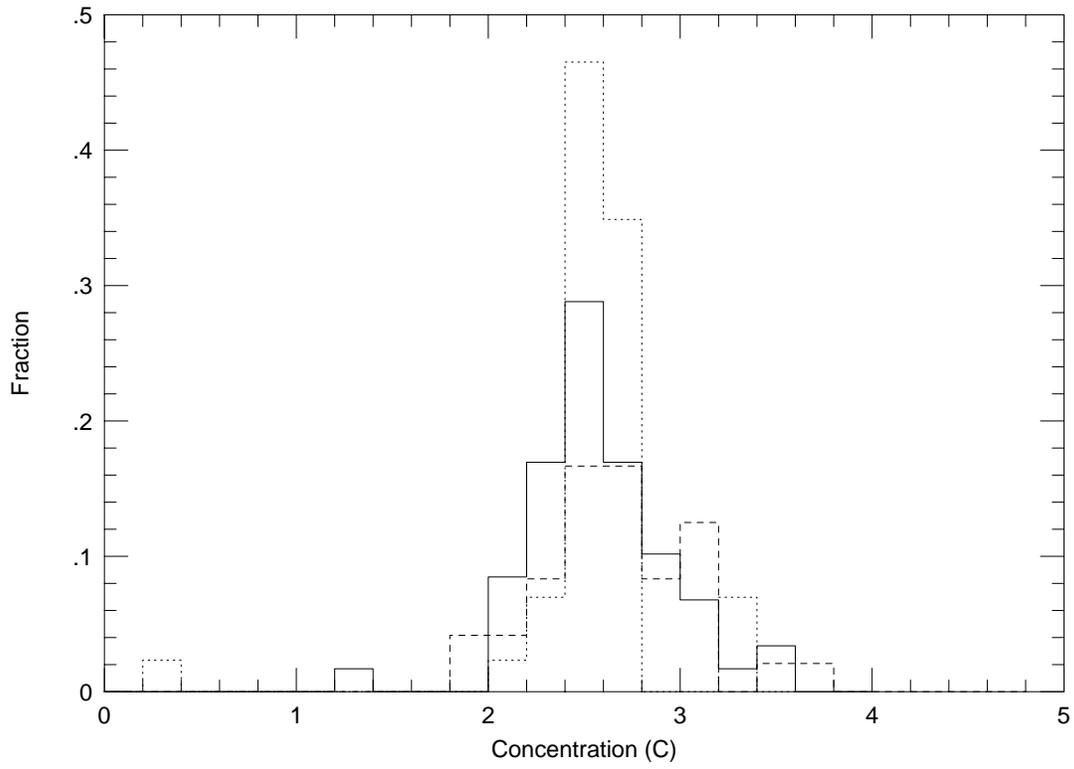}
\caption{Distribution of concentration (C) indices for our sample of resolved LAE components (solid line), systems (dashed line), and for a subsample
of stars from the GEMS images (dotted line).
\label{fig:conc}}
\end{figure}

Figure~\ref{fig:conc} displays the distribution of concentration indices, both for the 
individual LAE components and the multi-component systems.  From the figure, it appears
that our measurements are consistent with concentration indices $2 < C < 3.2$, which were 
found for high-redshift LAEs ($4.1 < z < 5.7$) by \citet{Pirzkal07} and \citet{Overzier08}.  
Our data also appear to agree with LBG measurements at $z \sim 3$, which have found
$1.6 < C < 4$, with a median value of $C = 2.69$ \citep{overzier10}.   However, to check
the robustness of our result, we examined the concentration indices for a subsample of 7 
resolved, high signal-to-noise ($S/N > 30$) LAEs in GOODS that were also present on the
shallower sGOODS frames.   For these objects, there was no systematic difference in the 
concentration values between sGOODS and GOODS ($\Delta C = -0.01$), but the standard deviation 
of the measurements was large ($\sigma_C = 0.37$).  
Moreover, when we determined the concentration indices for a set of MUSYC-ECDF-S field stars \citep{MartinStars} on the 
GEMS images, we found a median value of $C = 2.5$, with the vast majority of the stars 
having $2.4 < C < 2.6$.  Since the majority of the LAE components are also within this range, 
the implication is that the {\sl HST ACS\/} does not have the angular resolution to 
determine the concentration of high-redshift LAEs, even when they are formally identified as 
``resolved'' by {\small GALFIT}.  We note that previous LAE studies likely suffer from this
same limitation \citep{Pirzkal07,Overzier08}; without higher spatial resolution, the
concentrations of high$-z$ LAEs cannot be reliably determined.

\subsection{Ellipticity Distribution}
\label{subsec:ellipticity}
 
\begin{figure}[t]
\plotone{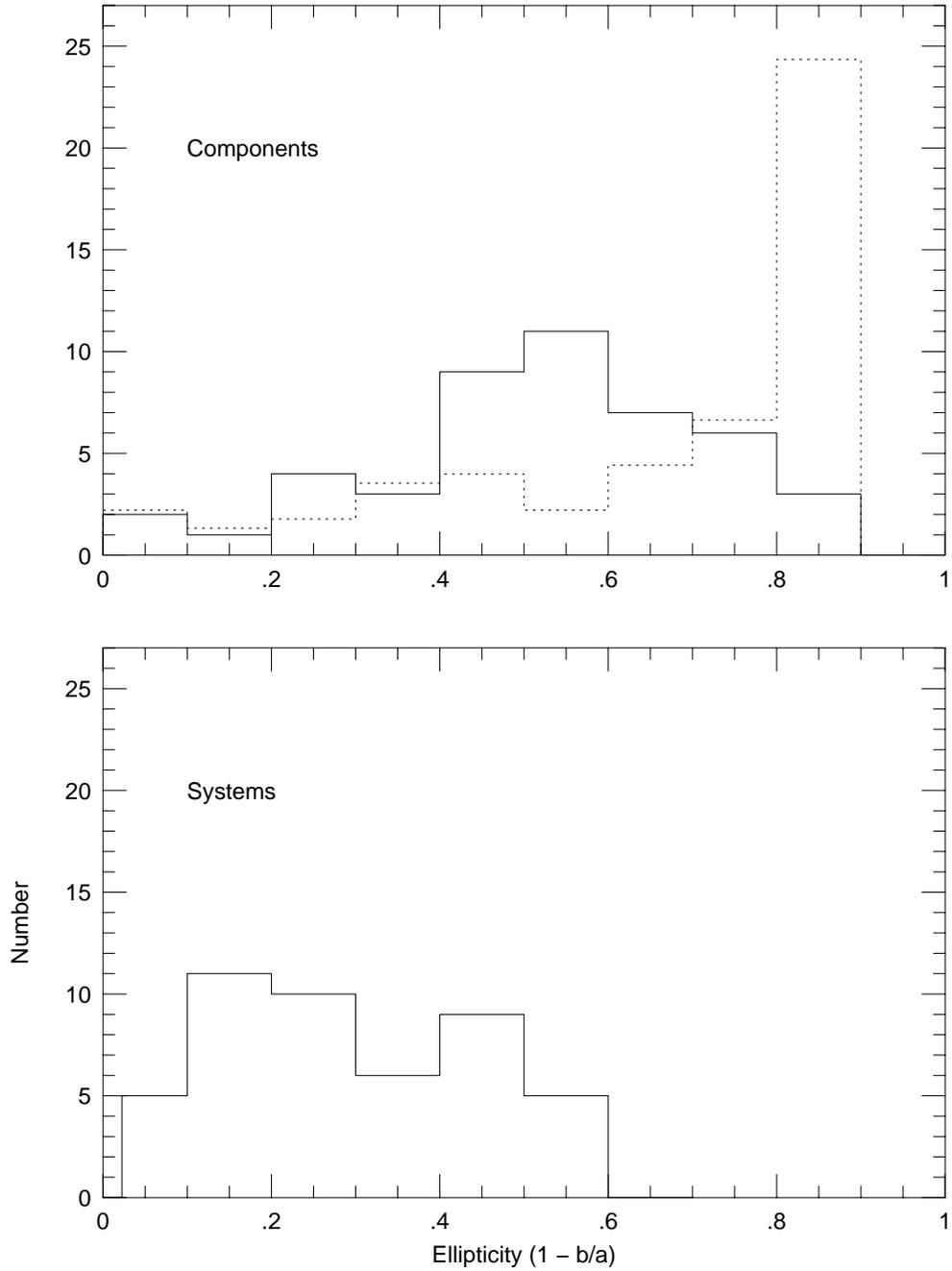}
\caption{Top panel: Distribution of ellipticities ($1 - b/a$) for our sample of resolved LAE components with $S/N > 30$ (shaded histogram) and for stars on
the GEMS images (solid black line).  Bottom panel:  Distribution of ellipticities for our sample of LAE systems with $S/N > 30$.}
\label{fig:ellipticities}
\end{figure}

As determined by measurements from the Automated Plate Machine (APM) and Sloan
Digital Sky Survey (SDSS),  the distribution of spiral galaxy ellipticities in the local
neighborhood is basically flat, with $\epsilon = 1 - b/a$ ranging from $\sim 0.2$ to 0.8 
\citep{lambas,padilla}.   We can compare this result to the ellipticities of resolved $z = 3.1$
LAEs with $S/N > 30$ by using the values produced by the PSF-convolved 2-dimensional 
elliptical models of {\small GALFIT\null.} This is done in Figure~\ref{fig:ellipticities}.  As the
figure illustrates, the distribution of ellipticities is not flat:  it is skewed towards higher
values and peaks at $\epsilon \sim 0.55$.  In a similar study using the same technique, 
\citet{Ravindranath06} found the same result for LBG galaxies at $z > 2.5$; in contrast,
star-forming galaxies at $z \sim 1.2$ had the same flat ellipticity distribution seen in the local universe. The distribution of ellipticities hints that at $z \sim 3$, the majority of LBGs and LAEs 
are elongated and prolate.  Some may even be similar to the ``chain" galaxies identified
in other surveys \citep[e.g.,][]{elmegreen}.
 

To test the robustness of this result, we again used the \citet{MartinStars}
sample of field stars to determine the ellipticity distribution of point sources on
the GEMS images.   Using IMEXAM in IRAF yielded values near zero (typically less than 0.05) as expected. 
However,  as Figure~\ref{fig:ellipticities} illustrates,  when fitting to a  S\'{e}rsic profile as was used for the galaxies,
{\small GALFIT}
finds that the vast majority of stars have high values for ellipticity ($\epsilon > 0.7$), rather than 
values near zero.     We have checked that the position angles reported by GALFIT for the stars are randomly
distributed, so we do not believe there is a systematic issue with the measured axis ratios.    We note that
\citet{benson07} found a similar result of unusually high ellipticities measured in a sample of bulges of SDSS galaxies 
using a different program \citep[GALACTICA;][]{benson02}.  \citet{benson07} attributed this discrepancy to not including
a needed a necessary component to the fit, in their case the presence of an elongated bar.   The 
 {\it HST/ACS\/} PSF has known thermal variations and aliasing produced by the drizzling process \citep{rhodes} which we 
 have not accounted for in our analysis which may be increasing the difficulty in measuring the ellipticity of an unresolved 
 source.
The main source of this odd result is the inherent difficulty associated with
fitting unresolved sources using PSF convolution.  
When the angular size of an object is close to that of a point-source a high-ellipticity model 
of its light profile will be as good a fit as a low-ellipticity model after PSF convolution.   Thus, 
the best-fit ellipticity as reported by {\small GALFIT} is reflecting the underlying noise in the image
and any errors in our model of the PSF.
Nevertheless, the experiment does
prove that the observed ellipticity distribution of resolved LAEs is not an artifact 
produced by the {\small GALFIT} program.  This confirms the results obtained by 
\citet{Ravindranath06} using Monte Carlo simulations: the
{\small GALFIT} algorithms are not responsible for the skewness seen in the figure.

In addition to using the {\small GALFIT} measurements of $\epsilon$, we also directly 
measured the second-order moments of each resolved, LAE's luminosity profile using the 
ellipse parameters of equation~\ref{eq:ellipse}.   These measurements do not take into
account the image PSF and therefore yield much smaller ellipticities ($\epsilon < 0.6$), 
particularly for single-component objects.  Our directly measured values for objects with 
$S/N > 30$ are shown in the bottom panel of Figure~\ref{fig:ellipticities}.    This distribution
is consistent with that found by \citet{Ferguson04} in a similar study of $z\sim4$
Lyman-break galaxies.

\subsection{S\'ersic Indices and Sizes}

\begin{figure}[t]
\plotone{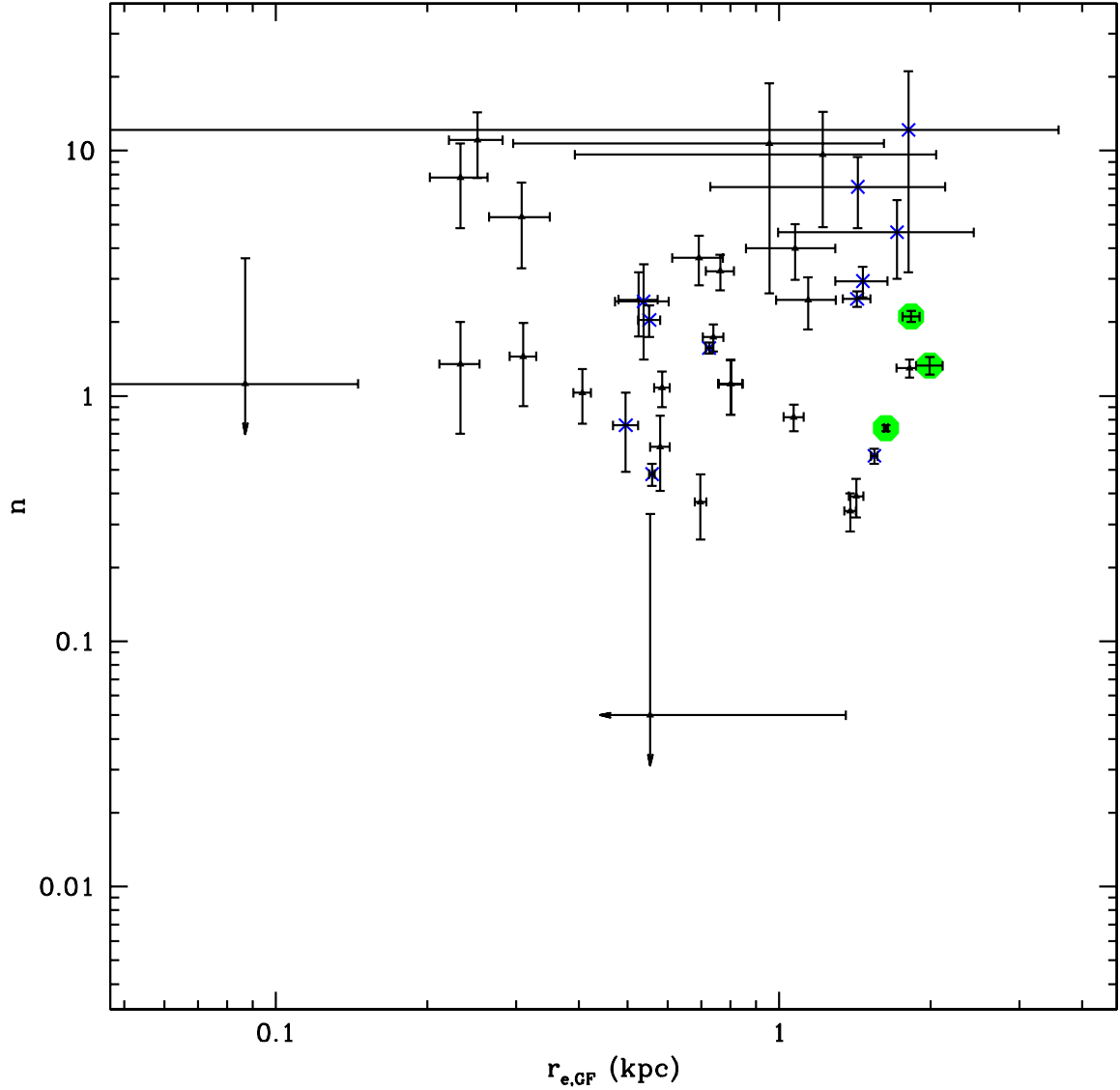}
\caption{The best-fit half-light radii
and S\'{e}rsic indices for our sample of resolved LAE components with $S/N > 30$ . Black triangles are from GEMS, blue crosses are from
GOODS, and green circles are from the HUDF.\null
There is a trend seen such that
components with $n\gtrsim 2$ have an excess of diffuse
emission at large radii, while components with $n \lesssim 2$ 
have extended or multi-compent light distributions.
\label{fig:Rn_GEMS}}
\end{figure}

Table~\ref{tab:Morph} gives the S\'erseic indices for all our resolved LAE components
with $S/N > 30$.  Of these 52 objects components analyzed, {\small GALFIT} converged
on a solution for 48 of them, while the results for two others (both components of LAE 12)
were discarded due to contamination from a bright extended source just outside the selection
circle.   These best-fit S\'{e}rsic indices, $n$, are plotted against the {\it intrinsic\/}
half-light radii, $r_{e,GF}$, in Figure~\ref{fig:Rn_GEMS}.   A cursory examination of
the figure reinforces the fact that most LAE components are extremely compact:
although UV emission from LAEs can have a range of half-light radii from $\la 0.3$~kpc (predominantly point sources, not plotted) to $\sim 2$~kpc, the majority are very small, with
$r_{e,GF}<1$~kpc.   This result is similar to that found for LAEs at $z=4.1$ to $5.7$ by \citet{Pirzkal07,Overzier08,taniguchi}.



\begin{figure}[t]
\plotone{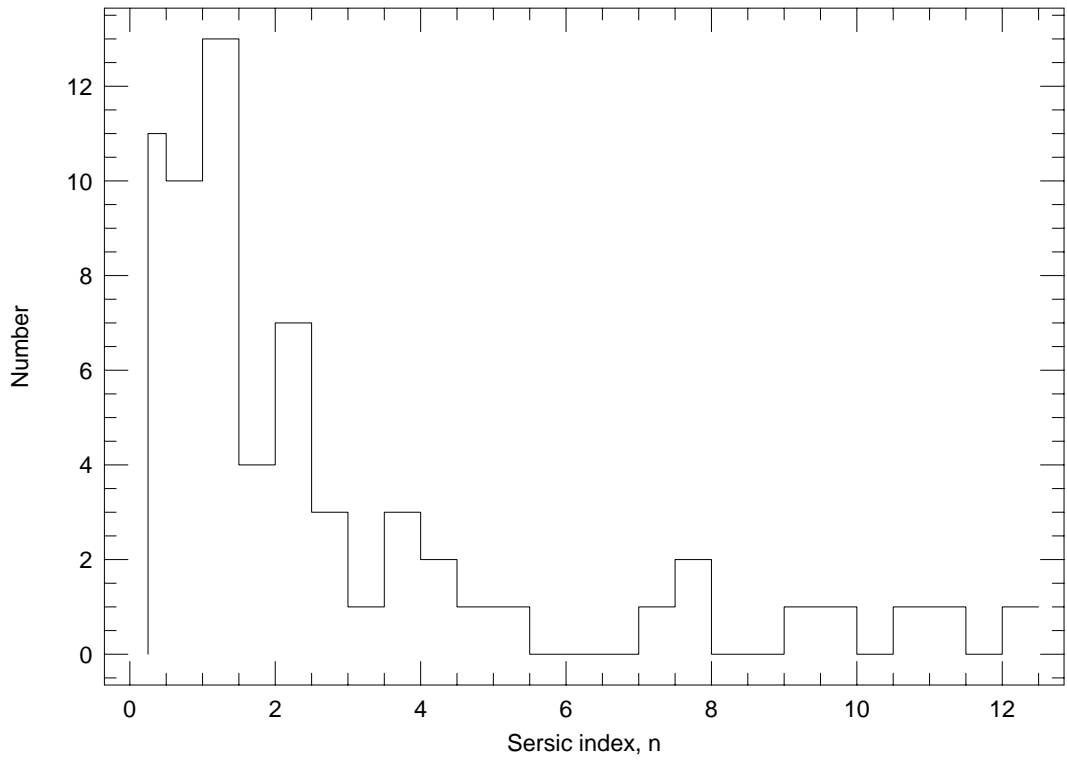}
\caption{Distribution of S\'ersic indices for our sample of resolved LAE components with $S/N>30$.  There is a broad distribution in $n$ with a peak at  $n \sim 1$.}
\label{fig:sersic}
\end{figure}

A histogram of 
the S\'ersic indices 
is shown in
Figure~\ref{fig:sersic}.  Although the distribution is broad, with values ranging from
$n \lesssim 1$ to $n \gtrsim 12$, the majority of LAEs have small S\'ersic
indices, and the peak of the distribution is at $n \sim 1.5$.  This is simlar to what is
seen for nearby galaxies from the Sloan Digital Sky Survey \citep{blanton} although our
distribution lacks the high-$n$ population associated with massive, red galaxies.  Moreover,
a visual inspection of the images shows a morphological transition at $n \sim 2$.  
Components with small $n$ generally have extended or split peaks in their luminosity profile, 
with relatively little diffuse emission beyond $2r_{e,GF}$.  While it is possible that these objects
be merging systems, multiple star-forming clumps in close proximity to each other can also
conspire to produce small values of $n$ \citep[e.g.,][]{rawat09}.  By contrast, LAE
components with $n \ga 2$ have high concentration cores, surrounded by diffuse emission 
that often extends to many $r_{e,GF}$.  
Interestingly, unlike in low-redshift spheroids, this 
diffuse emission is often amorphous or clumpy.  It is possible that these compact components
represent bulges in the act of formation, or, alternatively, isolated star-forming regions within a larger galaxy.

\subsubsection{Robustness of the S\'ersic Fits}

The uncertainties reported by GALFIT on the S\'{e}rsic index reported in 
Table~\ref{tab:Morph} are based upon the local curvature of the $\chi^2$ surfaces,
and therefore do not account of systematic errors in the analysis. In their
study of LBGs within the GOODS fields, \citet{Ravindranath06} simulated $50,000$
pure $n=1$ and $n=4$ systems with magnitudes $21 < V < 27$ and
half-light radii ($0\farcs01 < r_e < 5\farcs0$).  They found that for objects with
magnitudes ($25 < V < 26.6$) and sizes ($r_e < 0\farcs13$) considered
here, the {\small GALFIT} fitting procedure typically has a random uncertainty of 
$\sigma_n \sim 1.5$ for spheroids and $\sigma_n \sim 0.5$ for disks (see their
Figure~4).  Their simulations also demonstrate that the {\small GALFIT} algorithms
have a tendency to overestimate $n$ for disk systems ($\langle
n \rangle \sim 1.1$) and to underestimate $n$ in spheroids ($\langle n
\rangle \sim 3.83$).  \citet{Haussler} found similar results in their {\small GALFIT} sutdy
of high-redshift galaxies in GEMS\null.

An alternative way of investigating the systematic tendencies of 
{\small GALFIT} is to perform a direct comparison of independent observations of galaxies
taken to different depths.  Between HUDF, GOODS, and sGOODS, there are 6 LAEs with 
components that are resolved in more than one survey.  Table~\ref{tab:HUDFvALL}
lists these objects, along with the best-fit $V$-band magnitude, $V_{GF}$, half-light radii,
and S\'{e}rsic index for the brightest component of each object.  The table demonstrates that 
there is a slight tendency for high redshift LAEs to appear dimmer and smaller in 
shallower surveys, and for these surveys to overestimate $n$.  However, since the weighted 
mean difference between sGOODS and GOODS, $\Delta n_{GG} = 0.65 \pm 0.01$,
is significantly larger than that between HUDF and GOODS, $\Delta
n_{HG} = 0.215 \pm 0.001$, the data suggest that the GOODS imaging is deep enough to
allow us to distinguish point sources from more extended objects.

Figure~\ref{fig:HUDFPanels} displays the sGOODS, GOODS, and HUDF
images for the 3 LAE components in the HUDF field, and compares them
to their S\'ersic fits.   The brightest of the three sources, LAE 25 ($V_{GF} \sim 25$), 
is well fit by a S\'ersic disk ($n \sim 1$) with $r_{e,GF} \sim 1.5$~kpc.  (Additional
components are detected in the deeper surveys but  have $V_{GF}>27.5$ and are
indistinguishable from point sources.)   In contrast, the GOODS and sGOODS fits
fits for LAE $56$ and LAE $125$ (both with $V_{GF} \sim 26.5$) are highly uncertain, 
despite their large half-light radii ($\sim 1.8$~kpc).   This is due to the 
asymmetric associated with the diffuse emission surrounding the objects'
central cores: any attempt to fit these irregular features with a simple model will 
have difficulty.  In fact, in the sGOODS data, SExtractor divided the primary 
components of these objects in two, turning resolved LAEs into multiple discrete sources.  
Thus, the interpretation of LAEs 56 and 125 are not straightforward.

Three additional sources were resolved in both GOODS and sGOODS; these
are plotted in Figure~\ref{fig:GOODSPanels}.  Of the three, LAE $6$
has the brightest single component at $V_{GF}\sim 25.4$ and fits to
this source are fairly robust, with $r_{e,GF} \sim 0.6$~kpc and $n \la 0.5$. 
LAE 11, however, has two bright components within the selection circle
(only the results for the brightest component are given in Table~\ref{tab:HUDFvALL}) 
as well as some diffuse light, possibly associated with an interaction.  The
inconsistencies between the GOODS and sGOODS fits are primarily due to the 
additional (and likely spurious) detections in the sGOODS image.  Finally, 
for LAE 59, {\small GALFIT} produced single component models for both the
GOODS and the sGOODS data.  However, since the latter fit failed to generate
an error bar, we have removed the object from the table.


%

\section{Discussion}
\label{sec:discussion}

\begin{figure}[t]
\plotone{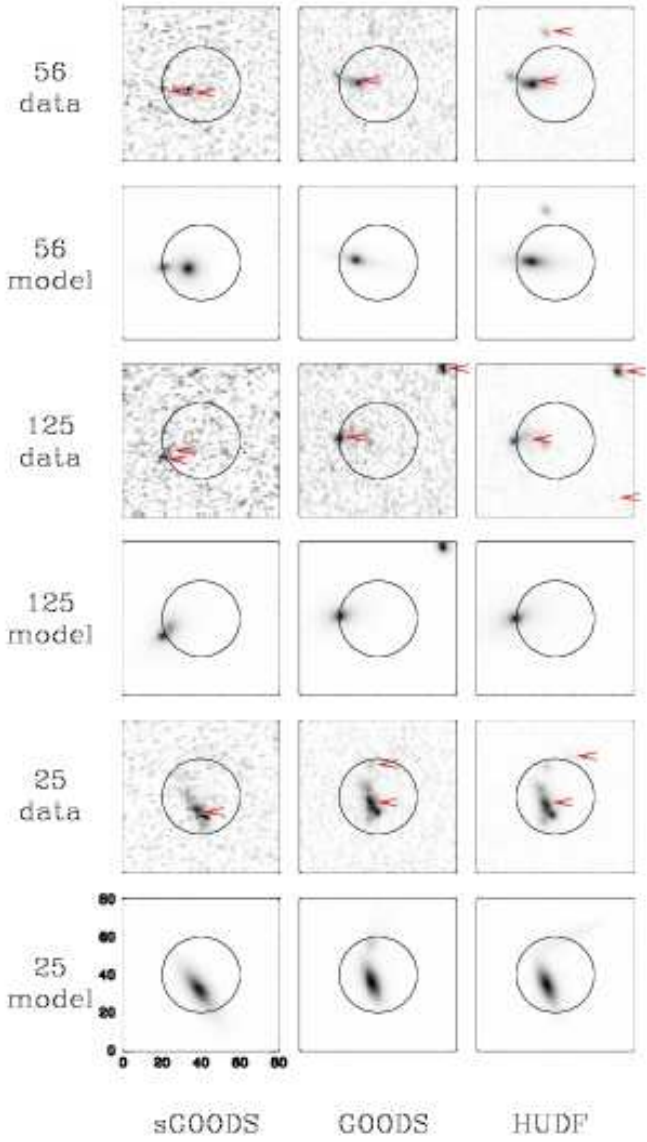}
\caption{The images and best-fit S\'{e}rsic profiles for the three LAEs with HUDF,
GOODS, and sGOODs data.  We mark all possible component  detections with red arrows 
in the data panels and draw the selection circle in black for all panels.  Cutouts are $2\farcs4 
\times 2\farcs4$ on a side.
\label{fig:HUDFPanels}}
\end{figure}
\begin{figure}[t]
\plotone{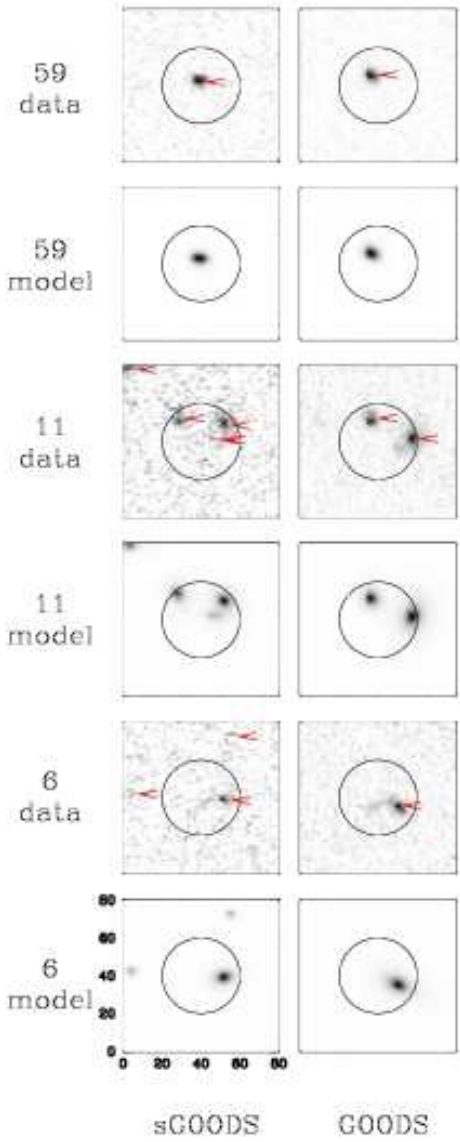}
\caption{Same as Figure~\ref{fig:HUDFPanels}, but for the three resolved LAEs
having only GOODS and sGOODS coverage.
\label{fig:GOODSPanels}}
\end{figure}

From Figures~\ref{fig:HUDFPanels} and \ref{fig:GOODSPanels} it is
clear that our sample of $z\sim3$ LAEs have 
a broad range of morphologies.  While most sources are 
small ($r_e < 2$ kpc) and concentrated ($2 < C < 4$), the rest-frame UV 
emission from a typical LAE is neither smooth nor symmetrical.   LAE
ellipticities are skewed towards higher values of $\epsilon$, in a manner
similar to that seen for LBGs at the same redshift \citep{Ravindranath06}.
This suggests the presence of  elongated, prolate structures,
which may be indicative of merging activity or ``chain" star formation.
Unfortunately, it is difficult to probe the concentration indices of the population:
even for the subset of LAEs that {\small GALFIT} calls resolved, the distribution
of $C$ values is similar to that measured for stars.  Next generation 
space-based imagers are needed to properly probe this parameter.


Our LAEs also possess a broad range of S\'ersic indices (see Figure~\ref{fig:sersic}).
As in the local universe \citep{blanton}, the distribution of $n$ values peaks near $\sim 1$ with a
broad distribution with values ranging from $n \lesssim 1$ and $n \gtrsim 12$.  Our population does not,
however,
have the high-$n$ population associated with luminous, red galaxies in the local universe.
Whether we are observing the formation of the Hubble sequence is not  clear, 
but there does appear to be a qualitative change in morphology at $n \sim
2$.  At greater $n$ values, LAE components appear more compact,
relative to their surrounding diffuse emission, than objects with $n \sim 1$.  However, 
these ``spheroid-like'' LAEs could also be the result of compact star-forming regions 
imbedded in the diffuse emission of a disk-like or irregular galaxy.  Deeper imaging is 
required to make this distinction.

  
If we classify LAEs with $n > 2.5$ as ``bulge-like" and LAEs with $2.5 \geq n > 0.8$ 
as ''disk-like" we find that 31\% of our sample of resolved LAE components with 
$S/N > 30$ fall into the former category, while 44\% belong to the latter.
In a sample of LBGs at the same redshift, \citet{Ravindranath06} found 27\% 
were bulge-like and 42\% were disk-like.   From the data, it appears that
the light profiles of LBGs and LAEs at this redshift have similar distributions.  

A subset of our LAEs consist of multiple components, and this can be interpreted as
evidence for merger activity.
When interacting systems are in close proximity, they may appear as 
as one single object with a small value of $n$ (see, for example, LAE $25$ in
Figure~\ref{fig:HUDFPanels}).  If we follow \citet{Ravindranath06} and
consider sources with $n < 0.8$ (i.e., with light profiles that are flatter than
exponential) to be merger candidates, then $\sim 24$\% of our sample would be classified 
as such.  This again is consistent with the 31\% value found by \citet{Ravindranath06} for 
LBGs at a similar redshift.  
We note that in a stacked sample of
$z  \sim5.7$ LAEs created by \citet{taniguchi}, $n \sim 0.7$ which may indicate a larger
fraction of merger candidates in LAE samples at higher redshift.   The sample of Taniguchi, however, covers a significantly
brighter range of Ly$\alpha$ luminosity ( $3 \times 10^{43}$  to $6.3 \times 10^{42}$ ergs~sec$^{-1}$) than our
sample at $z\sim3.1$ ($\sim 10^{43}$ to $1.3 \times 10^{42}$ ergs~sec$^{-1}$).   Ideally, such a comparison between
morphological
properties of LAE samples would be done by comparing samples at similar luminosities, but such samples do
not currently exist besides that presented here.

Of course, estimating merger rate using measured $n$ values fails to account for LAEs in 
which the interacting galaxies are at large separation.  In these cases,
SExtractor will consider the objects separate as components (see, for example, LAE $11$ in
Figure~\ref{fig:GOODSPanels}).  On the other hand, galaxies with
clumpy internal star formation can be mistaken for mergers when viewed
in the rest-frame UV\null.  Clumpy star-formation will lead also to low measured values
of $n$ \citep[e.g.,][]{rawat09} and hence an overestimate of the estimated merger rate.
To rule out this possibility, and to make progress towards determining 
the true LAE merger fraction, we would require deep $V$-band imaging
($m_{lim,5\sigma} \sim 30$) for a much larger sample of LAEs than is
covered by the HUDF\null.

Our morphological results for LAEs at $z\sim3$ 
are broadly similar to those found for continuum-selected LBG objects at the same redshift.
This consistency supports the contention of \citet{law} and \citet{pentericci} that 
that the presence of Ly$\alpha$ in emission has little bearing on rest-frame UV 
morphology of $z \sim 3$ star-forming galaxies.  
Comparable results were also found by \citet{taniguchi} at $z\sim6$, and may support
a scenario in which the observed differences between LAEs and LBGs are driven by 
variations in dust content and HI gas \citep{verhamme}.
On the other hand \citet{vanzella} found that dropout galaxies at $z\sim4$ with Ly$\alpha$-emission were significantly more concentrated than the rest of the sample.     Disentangling the effects of emission-line strength on morphology as a function of redshift will require 
substantially larger samples with deep high-resolution imaging at multiple redshifts, preferably tracking the same rest-frame
continuum wavelength free of contamination from the Ly$\alpha$ emission line.
 
Our results for the morphologies of $z \sim 3$ LAEs highlight the need for consistency 
between measurements made at different survey depths and the need for systematic checks 
to verify the robustness of derived parameters.  It also points out the importance of 
making measurements consistently over a wide range of redshifts.  
This paper represents the first attempt to measure morphological parameters of a
large sample of LAEs as previous work has concentrated primarily on measuring sizes
\citep{Venemans05,Bond09}, used small ($n \lesssim 12$) sample sizes \citep{Overzier08,Pirzkal07}, or
measured the profiles of a stacked sample due to low signal-to-noise \citep{taniguchi}.    The advent of larger samples of
LAEs from redshifts ranging from $z=2.1$ to $z>7$ should allow for meaningful comparisons across redshifts down to similar luminosity limits to be made.
Without such data,
it is impossible to form a clear picture of how the process of galaxy formation proceeds
over time.


The measurements presented here are limited by the fact that we are observing the
morphologies of these high-redshift galaxies in the rest-frame ultraviolet where individual
clumps of on-going star-formation may strongly affect the light distributions \citep[e.g.,][]{overzier10}.  Future work in the rest-frame optical (observed infrared) using WFC3 on
{\sl HST} should alleviate some of these concerns, and allow us the bulk of the LAE
stellar population.  A comparison of the two bandpasses will then enable us to 
examine how star formation proceeds in these objects.



\section{Acknowledgements}
\label{sec:ack}

Support for this work was provided by NASA through grants
HST-AR-10324.01-A \& HST-AR-11253.01-A from the Space Telescope Science Institute, which is
operated by AURA, Inc., under NASA contract NAS 5-26555 and by the
National Science Foundation under grants AST-0807885 \&  AST-0807570.   We thank 
Justin McKane for his assistance with the analysis.
Some of the data presented in this paper were obtained from the Multimission Archive at the Space Telescope Science Institute (MAST). STScI is operated by the Association of Universities for Research in Astronomy, Inc., under NASA contract NAS5-26555. Support for MAST for non-HST data is provided by the NASA Office of Space Science via grant NAG5-7584 and by other grants and contracts.

\clearpage
\bibliographystyle{apj}                       

\bibliography{apj-jour,ms}    


\begin{deluxetable}{lccccccccc}
\tablecaption{LAE Component Morphological Properties \label{tab:Morph}}
\rotate
\tablewidth{0pt}\tabletypesize{\scriptsize}
\tablehead{
\colhead{Object}
&\colhead{Component}
&\colhead{Survey}
&\colhead{$\alpha$}
&\colhead{$\delta$}
&\colhead{$V_{GF}$}
&\colhead{$r_{e,GF}$}
&\colhead{$n$}
&\colhead{$b/a$}
&\colhead{$C$}
\\
&
&
&
&
&
&\colhead{(kpc)}
&
&
}
\startdata
$25$ &$1$ &HUDF &$3$:$32$:$40.779$ &$-27$:$46$:$06.076$ &$25.11 \pm 0.01$ &$1.63 \pm 0.01$ &$0.74 \pm 0.02$ &$0.29 \pm 0.00$ & $2.60$  \\
$56$ &$1$ &HUDF &$3$:$32$:$34.324$ &$-27$:$47$:$59.559$ &$26.24 \pm 0.02$ &$1.83 \pm 0.07$ &$2.11 \pm 0.11$ &$0.25 \pm 0.01$  & $3.01$\\
$125$ &$1$ &HUDF &$3$:$32$:$39.017$ &$-27$:$46$:$22.287$ &$26.47 \pm 0.04$ &$1.99 \pm 0.12$ &$1.33 \pm 0.11$ &$0.40 \pm 0.02$ & $2.07$ \\
$4$ &$1$ &GOODS &$3$:$32$:$18.817$ &$-27$:$42$:$48.190$ &$25.25 \pm 0.02$ &$0.73 \pm 0.01$ &$1.57 \pm 0.08$ &$0.51 \pm 0.02$ & $2.55$\\
$ $ &$2$ &GOODS &$3$:$32$:$18.791$ &$-27$:$42$:$48.215$ &$27.42 \pm 0.32$ &$0.30 \pm 0.15$ &$7.97 \pm 11.08$ &$0.35 \pm 0.22$ &  \\
$ $ &$3$ &GOODS &$3$:$32$:$18.840$ &$-27$:$42$:$47.789$ &$27.60 \pm 0.33$ &$0.88 \pm 0.44$ &$2.52 \pm 2.10$ &$0.60 \pm 0.16$ &  \\
$ $ &$4$ &GOODS &$3$:$32$:$18.800$ &$-27$:$42$:$47.863$ &$27.62 \pm 0.11$ &$1.19 \pm 0.19$ &$1.08 \pm 0.50$ &$0.14 \pm 0.07$ &  \\
$6$ &$1$ &GOODS &$3$:$32$:$52.689$ &$-27$:$48$:$09.264$ &$25.37 \pm 0.01$ &$0.56 \pm 0.01$ &$0.48 \pm 0.05$ &$0.48 \pm 0.01$ & $2.32$\\
$11$ &$1$ &GOODS &$3$:$32$:$26.925$ &$-27$:$41$:$28.010$ &$25.41 \pm 0.04$ &$1.43 \pm 0.09$ &$2.49 \pm 0.18$ &$0.68 \pm 0.02$ &$2.85$ \\
$ $ &$2$ &GOODS &$3$:$32$:$26.968$ &$-27$:$41$:$27.723$ &$26.25 \pm 0.03$ &$0.55 \pm 0.03$ &$2.04 \pm 0.30$ &$0.77 \pm 0.05$  &$2.33$\\
$25$ &$1$ &GOODS &$3$:$32$:$40.783$ &$-27$:$46$:$06.057$ &$25.19 \pm 0.01$ &$1.55 \pm 0.02$ &$0.57 \pm 0.04$ &$0.30 \pm 0.01$ &$2.57$ \\
$35$ &$1$ &GOODS &$3$:$32$:$45.603$ &$-27$:$52$:$10.897$ &$27.02 \pm 0.03$ &$0.41 \pm 0.03$ &$1.29 \pm 0.46$ &$0.55 \pm 0.07$ & $2.52$\\
$41$ &$1$ &GOODS &$3$:$32$:$56.671$ &$-27$:$49$:$49.174$ &$26.63 \pm 0.05$ &$0.54 \pm 0.07$ &$2.43 \pm 1.02$ &$0.18 \pm 0.10$ &$2.67$ \\
$44$ &$1$ &GOODS &$3$:$32$:$15.790$ &$-27$:$44$:$09.902$ &$26.23 \pm 0.22$ &$1.43 \pm 0.70$ &$7.12 \pm 2.29$ &$0.60 \pm 0.07$ &$3.37$ \\
$ $ &$2$ &GOODS &$3$:$32$:$15.809$ &$-27$:$44$:$10.069$ &$26.91 \pm 0.09$ &$1.43 \pm 0.15$ &$1.11 \pm 0.19$ &$0.51 \pm 0.04$  &$2.25$\\
$56$ &$1$ &GOODS &$3$:$32$:$34.327$ &$-27$:$47$:$59.549$ &$26.50 \pm 0.37$ &$1.81 \pm 1.78$ &$12.16 \pm 8.96$ &$0.10 \pm 0.05$&$2.71$  \\
$59$ &$1$ &GOODS &$3$:$32$:$33.250$ &$-27$:$51$:$27.576$ &$25.80 \pm 0.07$ &$1.47 \pm 0.17$ &$2.94 \pm 0.42$ &$0.43 \pm 0.03$ &$3.19$ \\
$94$ &$1$ &GOODS &$3$:$32$:$09.335$ &$-27$:$43$:$54.177$ &$26.64 \pm 0.02$ &$0.50 \pm 0.03$ &$0.76 \pm 0.27$ &$0.59 \pm 0.05$ &$2.54$ \\
$125$ &$1$ &GOODS &$3$:$32$:$39.007$ &$-27$:$46$:$22.268$ &$26.40 \pm 0.21$ &$1.72 \pm 0.72$ &$4.65 \pm 1.65$ &$0.51 \pm 0.08$& \\
$8$ &$1$ &GEMS &$3$:$31$:$54.883$ &$-27$:$51$:$21.104$ &$25.44 \pm 0.01$ &$0.55 \pm 0.80$ &$0.05 \pm 0.28$ &$0.39 \pm 0.03$ & $2.44$\\
$9$ &$1$ &GEMS &$3$:$31$:$40.150$ &$-28$:$03$:$07.395$ &$25.17 \pm 0.01$ &$0.09 \pm 0.06$ &$1.12 \pm 2.52$ &$0.67 \pm 0.18$ &$2.58$ \\
$10$ &$1$ &GEMS &$3$:$33$:$22.453$ &$-27$:$46$:$36.909$ &$25.81 \pm 0.01$ &$0.41 \pm 0.02$ &$1.03 \pm 0.26$ &$0.22 \pm 0.04$&  \\
$13$ &$1$ &GEMS &$3$:$33$:$07.252$ &$-27$:$47$:$47.188$ &$25.58 \pm 0.04$ &$0.76 \pm 0.05$ &$3.23 \pm 0.53$ &$0.31 \pm 0.03$& $2.77$ \\
$15$ &$1$ &GEMS &$3$:$33$:$18.920$ &$-27$:$38$:$28.462$ &$26.53 \pm 0.03$ &$0.99 \pm 2.73$ &$0.03 \pm 0.37$ &$0.34 \pm 0.04$&$2.14$  \\
$17$ &$1$ &GEMS &$3$:$32$:$49.147$ &$-27$:$34$:$39.932$ &$25.59 \pm 0.03$ &$0.74 \pm 0.04$ &$1.74 \pm 0.22$ &$0.78 \pm 0.04$& $2.76$ \\
$18$ &$1$ &GEMS &$3$:$32$:$46.753$ &$-27$:$39$:$59.909$ &$26.68 \pm 0.03$ &$0.30 \pm 0.03$ &$1.39 \pm 0.73$ &$0.73 \pm 0.12$&$2.35$  \\
$19$ &$1$ &GEMS &$3$:$31$:$34.736$ &$-27$:$56$:$21.803$ &$25.03 \pm 0.09$ &$0.25 \pm 0.03$ &$11.06 \pm 3.31$ &$0.81 \pm 0.07$&$2.91$  \\
$20$ &$1$ &GEMS &$3$:$33$:$11.882$ &$-28$:$00$:$12.561$ &$26.40 \pm 0.07$ &$1.06 \pm 0.10$ &$1.01 \pm 0.17$ &$0.91 \pm 0.07$&$2.75$  \\
$ $ &$2$ &GEMS &$3$:$33$:$11.871$ &$-28$:$00$:$12.121$ &$26.58 \pm 0.03$ &$0.57 \pm 0.03$ &$0.90 \pm 0.35$ &$0.55 \pm 0.06$ &$2.53$ \\
$22$ &$1$ &GEMS &$3$:$31$:$51.634$ &$-27$:$58$:$32.617$ &$26.26 \pm 0.04$ &$0.31 \pm 0.04$ &$5.37 \pm 2.05$ &$0.16 \pm 0.08$&$2.83$  \\
$24$ &$1$ &GEMS &$3$:$31$:$53.212$ &$-27$:$57$:$08.156$ &$26.24 \pm 0.02$ &$0.23 \pm 0.02$ &$1.35 \pm 0.65$ &$0.46 \pm 0.10$&$2.61$  \\
$26$ &$1$ &GEMS &$3$:$31$:$51.560$ &$-27$:$46$:$47.005$ &$26.39 \pm 0.03$ &$1.02 \pm 0.04$ &$0.20 \pm 0.10$ &$0.58 \pm 0.03$&$2.05$  \\
$29$ &$1$ &GEMS &$3$:$31$:$47.795$ &$-27$:$45$:$03.300$ &$26.65 \pm 0.05$ &$0.41 \pm 0.09$ &$3.52 \pm 1.27$ &$0.37 \pm 0.11$&$1.20$  \\
$36$ &$1$ &GEMS &$3$:$32$:$18.925$ &$-27$:$38$:$40.183$ &$26.94 \pm 0.10$ &$0.65 \pm 0.10$ &$2.50 \pm 1.26$ &$0.48 \pm 0.11$&$3.53$  \\
$38$ &$1$ &GEMS &$3$:$31$:$50.358$ &$-27$:$59$:$10.099$ &$26.42 \pm 0.02$ &$0.56 \pm 2.04$ &$0.04 \pm 0.35$ &$0.26 \pm 0.08$&$2.25$  \\
$39$ &$1$ &GEMS &$3$:$31$:$30.522$ &$-27$:$47$:$29.637$ &$25.33 \pm 0.04$ &$1.82 \pm 0.10$ &$1.30 \pm 0.11$ &$0.39 \pm 0.02$&$2.84$  \\
$43$ &$1$ &GEMS &$3$:$33$:$07.314$ &$-27$:$54$:$38.985$ &$25.69 \pm 0.10$ &$0.23 \pm 0.03$ &$7.78 \pm 2.95$ &$0.97 \pm 0.10$&$2.70$  \\
$49$ &$1$ &GEMS &$3$:$31$:$42.357$ &$-27$:$58$:$07.834$ &$25.79 \pm 0.02$ &$1.38 \pm 0.04$ &$0.34 \pm 0.06$ &$0.38 \pm 0.01$&$2.34$  \\
$50$ &$1$ &GEMS &$3$:$31$:$52.829$ &$-27$:$45$:$18.631$ &$26.90 \pm 0.15$ &$0.76 \pm 0.05$ &$0.11 \pm 0.32$ &$0.12 \pm 0.12$&$2.16$  \\
$53$ &$1$ &GEMS &$3$:$32$:$15.130$ &$-27$:$38$:$53.937$ &$25.82 \pm 0.02$ &$1.42 \pm 0.05$ &$0.39 \pm 0.07$ &$0.48 \pm 0.02$&$2.24$  \\
$61$ &$1$ &GEMS &$3$:$33$:$09.435$ &$-27$:$45$:$50.102$ &$26.41 \pm 0.13$ &$1.07 \pm 0.23$ &$2.46 \pm 0.86$ &$0.47 \pm 0.08$&$2.15$  \\
$63$ &$1$ &GEMS &$3$:$32$:$51.916$ &$-27$:$42$:$12.234$ &$25.75 \pm 0.02$ &$0.59 \pm 0.02$ &$1.08 \pm 0.18$ &$0.54 \pm 0.03$&$2.63$  \\
$64$ &$1$ &GEMS &$3$:$31$:$59.830$ &$-27$:$49$:$46.413$ &$25.80 \pm 0.27$ &$1.22 \pm 0.83$ &$9.65 \pm 4.76$ &$0.70 \pm 0.11$&$2.58$  \\
$67$ &$1$ &GEMS &$3$:$32$:$51.768$ &$-27$:$37$:$33.540$ &$26.27 \pm 0.02$ &$0.31 \pm 0.02$ &$1.45 \pm 0.54$ &$0.42 \pm 0.06$&$2.06$  \\
$69$ &$1$ &GEMS &$3$:$33$:$25.356$ &$-28$:$02$:$46.519$ &$25.99 \pm 0.03$ &$1.07 \pm 0.05$ &$0.82 \pm 0.10$ &$0.53 \pm 0.03$&$3.42$  \\
$73$ &$1$ &GEMS &$3$:$32$:$57.403$ &$-27$:$55$:$19.055$ &$26.57 \pm 0.03$ &$0.31 \pm 0.03$ &$1.91 \pm 0.63$ &$0.70 \pm 0.09$&$3.72$  \\
$75$ &$1$ &GEMS &$3$:$32$:$59.267$ &$-27$:$41$:$14.734$ &$27.01 \pm 0.04$ &$0.49 \pm 0.04$ &$0.66 \pm 0.53$ &$0.10 \pm 0.17$&$2.50$  \\
$82$ &$1$ &GEMS &$3$:$31$:$47.775$ &$-27$:$42$:$16.311$ &$26.32 \pm 0.02$ &$0.58 \pm 0.03$ &$0.62 \pm 0.21$ &$0.55 \pm 0.04$&$2.37$  \\
$83$ &$1$ &GEMS &$3$:$31$:$38.669$ &$-27$:$45$:$43.551$ &$26.48 \pm 0.08$ &$0.85 \pm 0.10$ &$1.87 \pm 0.61$ &$0.49 \pm 0.07$&$2.45$  \\
$87$ &$1$ &GEMS &$3$:$33$:$05.025$ &$-27$:$43$:$37.295$ &$25.97 \pm 0.10$ &$1.08 \pm 0.22$ &$4.00 \pm 1.03$ &$0.49 \pm 0.06$&$3.55$  \\
$91$ &$1$ &GEMS &$3$:$31$:$58.802$ &$-27$:$49$:$28.753$ &$26.74 \pm 0.04$ &$0.88 \pm 4.64$ &$0.01 \pm 1.12$ &$0.38 \pm 0.07$&$2.43$  \\
$98$ &$1$ &GEMS &$3$:$31$:$26.618$ &$-27$:$44$:$02.175$ &$26.75 \pm 0.04$ &$1.04 \pm 0.08$ &$0.64 \pm 0.21$ &$0.53 \pm 0.05$&$2.18$  \\
$99$ &$1$ &GEMS &$3$:$31$:$40.245$ &$-27$:$45$:$26.702$ &$26.03 \pm 0.01$ &$0.70 \pm 0.02$ &$0.37 \pm 0.11$ &$0.41 \pm 0.03$&$2.82$  \\
$101$ &$1$ &GEMS &$3$:$33$:$07.747$ &$-27$:$38$:$19.332$ &$26.34 \pm 0.30$ &$0.96 \pm 0.66$ &$10.71 \pm 8.09$ &$0.22 \pm 0.08$&$2.58$  \\
$105$ &$1$ &GEMS &$3$:$33$:$12.401$ &$-27$:$45$:$24.291$ &$26.28 \pm 0.05$ &$0.53 \pm 0.05$ &$2.47 \pm 0.72$ &$0.53 \pm 0.07$&$3.13$  \\
$106$ &$1$ &GEMS &$3$:$32$:$21.284$ &$-27$:$36$:$21.398$ &$25.80 \pm 0.06$ &$0.69 \pm 0.08$ &$3.66 \pm 0.84$ &$0.71 \pm 0.06$&$3.59$  \\
$112$ &$1$ &GEMS &$3$:$32$:$43.954$ &$-27$:$37$:$16.846$ &$26.24 \pm 0.08$ &$1.14 \pm 0.16$ &$2.46 \pm 0.59$ &$0.32 \pm 0.05$&$2.35$  \\
$121$ &$1$ &GEMS &$3$:$33$:$23.709$ &$-27$:$44$:$09.126$ &$27.04 \pm 0.00$ &$0.50 \pm 0.00$ &$0.61 \pm 0.00$ &$0.44 \pm 0.00$&$2.30$  \\
$122$ &$1$ &GEMS &$3$:$32$:$20.467$ &$-27$:$35$:$01.616$ &$26.45 \pm 0.07$ &$0.46 \pm 0.06$ &$3.61 \pm 1.42$ &$0.45 \pm 0.08$&$3.04$  \\
$124$ &$1$ &GEMS &$3$:$31$:$42.924$ &$-28$:$03$:$07.835$ &$26.31 \pm 0.03$ &$0.80 \pm 0.04$ &$1.12 \pm 0.28$ &$0.25 \pm 0.04$&$2.86$  \\
$133$ &$1$ &GEMS &$3$:$31$:$42.951$ &$-27$:$45$:$06.568$ &$26.97 \pm 0.45$ &$1.74 \pm 2.04$ &$9.22 \pm 8.60$ &$0.10 \pm 0.08$&$2.53$  \\
$149$ &$1$ &GEMS &$3$:$33$:$07.026$ &$-27$:$37$:$53.498$ &$26.37 \pm 0.17$ &$1.43 \pm 0.49$ &$4.44 \pm 1.71$ &$0.30 \pm 0.06$&$2.67$  \\
$156$ &$1$ &GEMS &$3$:$33$:$00.602$ &$-28$:$00$:$06.503$ &$26.88 \pm 0.05$ &$0.80 \pm 0.06$ &$0.98 \pm 0.37$ &$0.36 \pm 0.07$&$2.59$  \\
$162$ &$1$ &GEMS &$3$:$33$:$15.186$ &$-27$:$54$:$01.859$ &$26.82 \pm 0.03$ &$0.64 \pm 0.03$ &$0.08 \pm 0.36$ &$0.48 \pm 0.04$&  \\
\enddata

\end{deluxetable}

\begin{deluxetable}{lccccccc}
\tablecaption{LAE System Morphological Properties\label{tab:systems}}
\tablewidth{0pt}\tabletypesize{\scriptsize}
\tablehead{
&&&&\colhead{$V^{\rm PHOT}$} 
&\colhead{$r_e^{\rm PHOT}$\tablenotemark{c}} & & \\
\colhead{Number\tablenotemark{a}}
&\colhead{Survey}
&\colhead{$\alpha$\tablenotemark{b}}
&\colhead{$\delta$\tablenotemark{b}}
&\colhead{(AB mags)}
&\colhead{(kpc)}
&\colhead{$b/a$\tablenotemark{d}}
&\colhead{$C$}
}
\startdata
25  &HUDF  &$3$:$32$:$40.785$ &$-27$:$46$:$06.035$ &$25.04 \pm 0.01$ &1.48   &0.57 &2.60 \\
56  &HUDF  &$3$:$32$:$34.329$ &$-27$:$47$:$59.543$ &$26.30 \pm 0.02$ &1.39   &0.55 &3.01 \\
125 &HUDF  &$3$:$32$:$39.012$ &$-27$:$46$:$22.307$ &$26.47 \pm 0.02$ &1.68   &0.67 &2.07 \\
4   &GOODS &$3$:$32$:$18.813$ &$-27$:$42$:$48.103$ &$24.89 \pm 0.03$ &1.49   &0.80 &3.14 \\
6   &GOODS &$3$:$32$:$52.690$ &$-27$:$48$:$09.288$ &$25.38 \pm 0.03$ &0.71   &0.86 &2.32 \\
11  &GOODS &$3$:$32$:$26.938$ &$-27$:$41$:$27.935$ &$25.16 \pm 0.03$ &2.56   &0.41 &\dots\\
25  &GOODS &$3$:$32$:$40.785$ &$-27$:$46$:$05.999$ &$25.04 \pm 0.03$ &1.52   &0.54 &2.51 \\
35  &GOODS &$3$:$32$:$45.604$ &$-27$:$52$:$10.913$ &$26.81 \pm 0.11$ &0.82   &0.89 &2.52 \\
41  &GOODS &$3$:$32$:$56.672$ &$-27$:$49$:$49.202$ &$26.78 \pm 0.20$ &0.68   &0.58 &2.67 \\
44  &GOODS &$3$:$32$:$15.799$ &$-27$:$44$:$09.992$ &$26.01 \pm 0.05$ &1.56   &0.61 &1.97 \\
55  &GOODS &$3$:$32$:$59.976$ &$-27$:$50$:$26.308$ &$26.37 \pm 0.18$ &1.28   &0.47 &2.29 \\
56  &GOODS &$3$:$32$:$34.331$ &$-27$:$47$:$59.550$ &$26.44 \pm 0.11$ &1.46   &0.54 &2.71 \\
59  &GOODS &$3$:$32$:$33.254$ &$-27$:$51$:$27.590$ &$25.86 \pm 0.06$ &1.33   &0.68 &3.19 \\
66  &GOODS &$3$:$32$:$48.528$ &$-27$:$56$:$05.374$ &$26.66 \pm 0.20$ &1.93   &0.73 &3.16 \\
85  &GOODS &$3$:$32$:$59.824$ &$-27$:$53$:$05.766$ &$26.62 \pm 0.13$ &0.74   &0.90 &2.54 \\
90  &GOODS &$3$:$32$:$14.574$ &$-27$:$45$:$52.420$ &$26.90 \pm 0.16$ &0.95   &0.71 &3.14 \\
94  &GOODS &$3$:$32$:$09.336$ &$-27$:$43$:$54.192$ &$26.63 \pm 0.13$ &0.72   &0.92 &2.53 \\
125 &GOODS &$3$:$32$:$39.016$ &$-27$:$46$:$22.307$ &$26.32 \pm 0.09$ &1.58   &0.75 &\dots\\
5   &GEMS  &$3$:$32$:$47.517$ &$-27$:$58$:$07.705$ &$24.95 \pm 0.03$ &1.14   &0.52 &2.34 \\
8   &GEMS  &$3$:$31$:$54.885$ &$-27$:$51$:$21.114$ &$25.32 \pm 0.04$ &0.73   &0.88 &2.50 \\
9   &GEMS  &$3$:$31$:$40.157$ &$-28$:$03$:$07.405$ &$25.06 \pm 0.03$ &0.58   &0.93 &2.58 \\
10  &GEMS  &$3$:$33$:$22.442$ &$-27$:$46$:$36.851$ &$25.12 \pm 0.04$ &2.17   &0.53 &\dots\\
12  &GEMS  &$3$:$32$:$33.846$ &$-27$:$36$:$35.118$ &$25.26 \pm 0.04$ &2.30   &0.59 &\dots\\
13  &GEMS  &$3$:$33$:$07.253$ &$-27$:$47$:$47.177$ &$25.72 \pm 0.07$ &0.82   &0.78 &2.77 \\
15  &GEMS  &$3$:$33$:$18.916$ &$-27$:$38$:$28.468$ &$25.75 \pm 0.07$ &1.99   &0.44 &\dots\\
17  &GEMS  &$3$:$32$:$49.148$ &$-27$:$34$:$39.972$ &$25.55 \pm 0.06$ &1.03   &0.84 &2.76 \\
18  &GEMS  &$3$:$32$:$46.753$ &$-27$:$39$:$59.915$ &$26.72 \pm 0.16$ &0.63   &0.93 &2.35 \\
19  &GEMS  &$3$:$31$:$34.738$ &$-27$:$56$:$21.818$ &$25.08 \pm 0.04$ &0.81   &0.90 &2.91 \\
20  &GEMS  &$3$:$33$:$11.879$ &$-28$:$00$:$12.380$ &$25.82 \pm 0.07$ &1.93   &0.42 &\dots\\
22  &GEMS  &$3$:$31$:$51.636$ &$-27$:$58$:$32.646$ &$26.18 \pm 0.10$ &0.78   &0.74 &2.83 \\
24  &GEMS  &$3$:$31$:$53.213$ &$-27$:$57$:$08.161$ &$26.18 \pm 0.09$ &0.61   &0.88 &2.61 \\
26  &GEMS  &$3$:$31$:$51.562$ &$-27$:$46$:$47.014$ &$26.16 \pm 0.10$ &1.17   &0.79 &2.05 \\
29  &GEMS  &$3$:$31$:$47.801$ &$-27$:$45$:$03.384$ &$26.35 \pm 0.12$ &1.23   &0.67 &2.89 \\
36  &GEMS  &$3$:$32$:$18.926$ &$-27$:$38$:$40.196$ &$26.99 \pm 0.21$ &0.81   &0.72 &3.53 \\
38  &GEMS  &$3$:$31$:$50.370$ &$-27$:$59$:$10.122$ &$26.52 \pm 0.13$ &0.60   &0.87 &2.25 \\
39  &GEMS  &$3$:$31$:$30.524$ &$-27$:$47$:$29.648$ &$25.33 \pm 0.04$ &1.58   &0.74 &2.84 \\
41  &GEMS  &$3$:$32$:$56.672$ &$-27$:$49$:$49.242$ &$26.55 \pm 0.14$ &0.94   &0.75 &3.07 \\
43  &GEMS  &$3$:$33$:$07.315$ &$-27$:$54$:$38.988$ &$25.78 \pm 0.07$ &0.74   &0.90 &2.70 \\
49  &GEMS  &$3$:$31$:$42.359$ &$-27$:$58$:$07.856$ &$25.69 \pm 0.06$ &1.26   &0.66 &2.34 \\
52  &GEMS  &$3$:$33$:$21.363$ &$-27$:$38$:$36.337$ &$26.91 \pm 0.19$ &0.55   &0.87 &2.69 \\
53  &GEMS  &$3$:$32$:$15.131$ &$-27$:$38$:$53.948$ &$25.82 \pm 0.07$ &1.25   &0.79 &\dots\\
55  &GEMS  &$3$:$32$:$59.982$ &$-27$:$50$:$26.369$ &$26.37 \pm 0.12$ &1.64   &0.75 &2.50 \\
58  &GEMS  &$3$:$33$:$06.943$ &$-27$:$42$:$27.853$ &$26.61 \pm 0.14$ &1.03   &0.79 &3.59 \\
61  &GEMS  &$3$:$33$:$09.422$ &$-27$:$45$:$50.112$ &$26.08 \pm 0.07$ &1.79   &0.90 &\dots\\
63  &GEMS  &$3$:$32$:$51.918$ &$-27$:$42$:$12.247$ &$25.66 \pm 0.06$ &0.81   &0.85 &2.63 \\
64  &GEMS  &$3$:$31$:$59.831$ &$-27$:$49$:$46.427$ &$25.99 \pm 0.08$ &1.13   &0.92 &2.57 \\
67  &GEMS  &$3$:$32$:$51.769$ &$-27$:$37$:$33.550$ &$26.45 \pm 0.12$ &0.55   &0.88 &2.44 \\
68  &GEMS  &$3$:$32$:$58.140$ &$-27$:$48$:$04.878$ &$26.70 \pm 0.16$ &1.34   &0.28 &2.17 \\
69  &GEMS  &$3$:$33$:$25.356$ &$-28$:$02$:$46.532$ &$26.12 \pm 0.09$ &0.99   &0.76 &2.54 \\
73  &GEMS  &$3$:$32$:$57.404$ &$-27$:$55$:$19.074$ &$26.31 \pm 0.11$ &0.85   &0.83 &3.78 \\
74  &GEMS  &$3$:$33$:$18.588$ &$-27$:$45$:$42.620$ &$26.67 \pm 0.13$ &0.74   &0.89 &2.40 \\
75  &GEMS  &$3$:$32$:$59.268$ &$-27$:$41$:$14.752$ &$26.87 \pm 0.19$ &0.77   &0.69 &2.50 \\
79  &GEMS  &$3$:$31$:$58.027$ &$-27$:$47$:$30.332$ &$26.26 \pm 0.09$ &1.70   &0.43 &2.92 \\
82  &GEMS  &$3$:$31$:$47.776$ &$-27$:$42$:$16.326$ &$26.43 \pm 0.12$ &0.67   &0.87 &2.38 \\
83  &GEMS  &$3$:$31$:$38.670$ &$-27$:$45$:$43.589$ &$26.52 \pm 0.11$ &0.96   &0.73 &2.52 \\
87  &GEMS  &$3$:$33$:$05.026$ &$-27$:$43$:$37.308$ &$25.98 \pm 0.08$ &1.26   &0.80 &3.55 \\
89  &GEMS  &$3$:$33$:$12.016$ &$-27$:$58$:$39.929$ &$26.66 \pm 0.16$ &1.38   &0.12 &2.91 \\
91  &GEMS  &$3$:$31$:$58.803$ &$-27$:$49$:$28.765$ &$26.91 \pm 0.19$ &0.63   &0.78 &2.43 \\
92  &GEMS  &$3$:$33$:$03.319$ &$-27$:$41$:$39.041$ &$26.98 \pm 0.21$ &2.32   &0.41 &\dots\\
98  &GEMS  &$3$:$31$:$26.621$ &$-27$:$44$:$02.177$ &$26.55 \pm 0.14$ &1.24   &0.73 &2.12 \\
99  &GEMS  &$3$:$31$:$40.241$ &$-27$:$45$:$26.827$ &$25.78 \pm 0.07$ &1.43   &0.91 &1.98 \\
101 &GEMS  &$3$:$33$:$07.750$ &$-27$:$38$:$19.356$ &$26.61 \pm 0.15$ &0.86   &0.52 &2.58 \\
105 &GEMS  &$3$:$33$:$12.403$ &$-27$:$45$:$24.307$ &$26.28 \pm 0.11$ &0.81   &0.88 &3.13 \\
106 &GEMS  &$3$:$32$:$21.285$ &$-27$:$36$:$21.341$ &$25.62 \pm 0.06$ &1.31   &0.81 &3.09 \\
112 &GEMS  &$3$:$32$:$44.073$ &$-27$:$37$:$17.810$ &$26.41 \pm 0.12$ &0.96   &0.59 &2.35 \\
113 &GEMS  &$3$:$31$:$35.944$ &$-27$:$50$:$52.915$ &$26.57 \pm 0.14$ &1.75   &0.79 &3.20 \\
121 &GEMS  &$3$:$33$:$23.709$ &$-27$:$44$:$09.139$ &$26.97 \pm 0.21$ &0.71   &0.82 &2.30 \\
122 &GEMS  &$3$:$32$:$20.465$ &$-27$:$35$:$01.630$ &$26.48 \pm 0.13$ &0.75   &0.98 &3.03 \\
124 &GEMS  &$3$:$31$:$42.925$ &$-28$:$03$:$07.798$ &$26.01 \pm 0.08$ &1.22   &0.64 &\dots\\
126 &GEMS  &$3$:$31$:$44.374$ &$-27$:$50$:$57.700$ &$26.29 \pm 0.11$ &0.91   &0.91 &3.67 \\
127 &GEMS  &$3$:$33$:$02.820$ &$-27$:$57$:$17.507$ &$26.94 \pm 0.20$ &0.74   &0.86 &2.64 \\
149 &GEMS  &$3$:$33$:$07.027$ &$-27$:$37$:$53.512$ &$26.37 \pm 0.12$ &1.27   &0.64 &2.67 \\
152 &GEMS  &$3$:$33$:$29.304$ &$-27$:$36$:$41.785$ &$25.93 \pm 0.08$ &2.25   &0.67 &2.51 \\
156 &GEMS  &$3$:$33$:$00.604$ &$-28$:$00$:$06.538$ &$26.88 \pm 0.18$ &0.85   &0.73 &2.59 \\
157 &GEMS  &$3$:$33$:$28.389$ &$-27$:$45$:$09.634$ &$26.84 \pm 0.19$ &1.71   &0.72 &2.28 \\
162 &GEMS  &$3$:$33$:$15.184$ &$-27$:$54$:$01.638$ &$25.74 \pm 0.07$ &2.31   &0.54 &\dots\\
\enddata
\tablenotetext{a}{Index from Table 2 of Gronwall et al.\ 2007}
\tablenotetext{b}{Position of ACS centroid (set to ground-based position when there are no SExtractor detections)}
\tablenotetext{c}{Half-light radius computed by {\tt PHOT} (not reported for LAEs without SExtractor detections)}
\tablenotetext{d}{Isophotal axis ratio about ACS centroid}

\end{deluxetable}

\begin{deluxetable}{lccccccccc}
\tablecaption{Morphological Properties vs. Depth for LAEs with Multiple Coverage\label{tab:HUDFvALL}}
\rotate
\tablewidth{0pt}\tabletypesize{\scriptsize}
\tablehead{
\colhead{Number}
&\multicolumn{3}{|c|}{$V_{GF}$}
&\multicolumn{3}{|c|}{$r_{e,GF}$}
&\multicolumn{3}{|c|}{$n$}
\\
&\colhead{HUDF}
&\colhead{GOODS}
&\colhead{sGOODS}
&\colhead{HUDF}
&\colhead{GOODS}
&\colhead{sGOODS}
&\colhead{HUDF}
&\colhead{GOODS}
&\colhead{sGOODS}
\\
&(AB mags)
&(AB mags)
&(AB mags)
&(kpc)
&(kpc)
&(kpc)
&
&
&
}

\startdata
GLAE$6$ &$--$ &$25.37 \pm 0.01$ &$25.39 \pm 0.07$ &$--$ &$0.56 \pm 0.01$ &$0.58 \pm 0.02$ &$--$ &$0.48 \pm 0.05$ &$0.10 \pm 0.14$  \\
GLAE$11$ &$--$ &$25.41 \pm 0.04$ &$25.94 \pm 0.20$ &$--$ &$1.43 \pm 0.09$ &$0.94 \pm 0.31$ &$--$ &$2.49 \pm 0.18$ &$4.22 \pm 1.47$  \\
GLAE$25$ &$25.11 \pm 0.01$ &$25.19 \pm 0.01$ &$24.92 \pm 0.03$ &$1.63 \pm 0.01$ &$1.55 \pm 0.02$ &$2.01 \pm 0.10$ &$0.74 \pm 0.02$ &$0.57 \pm 0.04$ &$1.13 \pm 0.09$  \\
GLAE$56$ &$26.24 \pm 0.02$ &$26.50 \pm 0.37$ &$26.68 \pm 0.15$ &$1.83 \pm 0.07$ &$1.81 \pm 1.78$ &$0.71 \pm 0.15$ &$2.11 \pm 0.11$ &$12.16 \pm 8.96$ &$1.85 \pm 1.09$  \\
GLAE$125$ &$26.47 \pm 0.04$ &$26.40 \pm 0.21$ &$26.90 \pm 0.67$ &$1.99 \pm 0.12$ &$1.71 \pm 0.72$ &$1.74 \pm 1.79$ &$1.33 \pm 0.11$ &$4.65 \pm 1.65$ &$3.22 \pm 2.66$  \\
\enddata

\end{deluxetable}

\end{document}